\documentclass{article}

\usepackage{arxiv}

\usepackage[utf8]{inputenc}
\usepackage[T1]{fontenc}
\usepackage{hyperref}
\usepackage{url}
\usepackage{booktabs}
\usepackage{amsfonts}
\usepackage{amsmath}
\usepackage{nicefrac}
\usepackage{microtype}
\usepackage{graphicx}
\graphicspath{ {./images/} }

\title{MoE-GraphSAGE-Based Integrated Evaluation of Transient Rotor Angle and Voltage Stability in Power Systems}

\author{
 Kunyu Zhang$^1$ \\
  Arizona State University\\
  \And
 Guang Yang$^2$ \\
  Xinjiang University\\
  \And
 Fashun Shi$^{3*}$ \\
  Tsinghua University\\
  \texttt{shi\_fashun@tsinghua.edu.cn} \\
  \And
 Shaoying He$^4$ \\
  Zhengzhou Weiguang Semiconductor Co., Ltd.\\
  \And
 Yuchi Zhang$^5$ \\
  City University of Hong Kong\\
}

\date{}

\begin{document}
\maketitle

\renewcommand{\thefootnote}{}
\footnotetext{\textcopyright{} 20XX IEEE. Personal use of this material is permitted. Permission from IEEE must be obtained for all other uses, in any current or future media, including reprinting/republishing this material for advertising or promotional purposes, creating new collective works, for resale or redistribution to servers or lists, or reuse of any copyrighted component of this work in other works.}
\renewcommand{\thefootnote}{\arabic{footnote}}

\begin{abstract}
The large-scale integration of renewable energy and power electronic devices has increased the complexity of power system stability, making transient stability assessment more challenging. Conventional methods are limited in both accuracy and computational efficiency. To address these challenges, this paper proposes MoE-GraphSAGE, a graph neural network framework based on the MoE for unified TAS and TVS assessment. The framework leverages GraphSAGE to capture the power grid's spatiotemporal topological features and employs multi-expert networks with a gating mechanism to model distinct instability modes jointly. Experimental results on the IEEE 39-bus system demonstrate that MoE-GraphSAGE achieves superior accuracy and efficiency, offering an effective solution for online multi-task transient stability assessment in complex power systems.
\end{abstract}


\section{Introduction}

Driven by the policy of ``double carbon'' goals~\cite{ruan2022}, a large-scale influx of new types of loads into the power grid has led to frequent changes in its topology and operating modes~\cite{kang2017}, posing severe challenges to the safe and stable operation of the system~\cite{li2022}. Conventional methods like time-domain simulation~\cite{zhu2016} and transient energy functions~\cite{zhang2019} constrained by limited computational efficiency. With the rapid increase in power system scale and complexity, these constraints prevent traditional methods from satisfying the requirements of current online applications. Artificial intelligence offers new perspectives for addressing this challenge.

Currently, some scholars have applied graph neural networks to transient stability~\cite{zhong2022,luo2021}, capturing topological dependencies but typically focusing on single stability types and neglecting the coupling between rotor angle and voltage stability.

To address this limitation, some studies have explored integrated evaluation of power angle and voltage. Reference~\cite{li2025} applies multi-task evaluation with dynamic heterogeneous graphs and the maximum Lyapunov exponent (MLE), but does not fully capture time-varying topological changes during disturbances. Reference~\cite{shi2023} employs parallel 1D-CNNs with attention mechanisms to extract voltage and power angle features, improving stability margin accuracy, yet its limited input features restrict capturing complex dynamic couplings.

In summary, existing studies lack a method that balances accuracy and efficiency for integrated assessment of power angle and voltage instability. To address this, we propose a gated mixture-of-experts (MoE) graph neural network for simultaneous evaluation of transient power angle and voltage stability. GraphSAGE captures spatiotemporal topological features of the grid, while the gating mechanism dynamically assigns tasks to expert subnetworks, facilitating specialized learning for diverse stability patterns~\cite{li2026ha}. A multi-task learning framework jointly assesses power angle and voltage stability, preventing information loss from separate evaluation. Load-balancing regularization further improves inter-expert coordination and model generalization.

\section{Multi Stability Assessment Model Based on MoE-GraphSAGE}

\begin{figure}[htbp]
  \centering
  \includegraphics[width=0.9\linewidth]{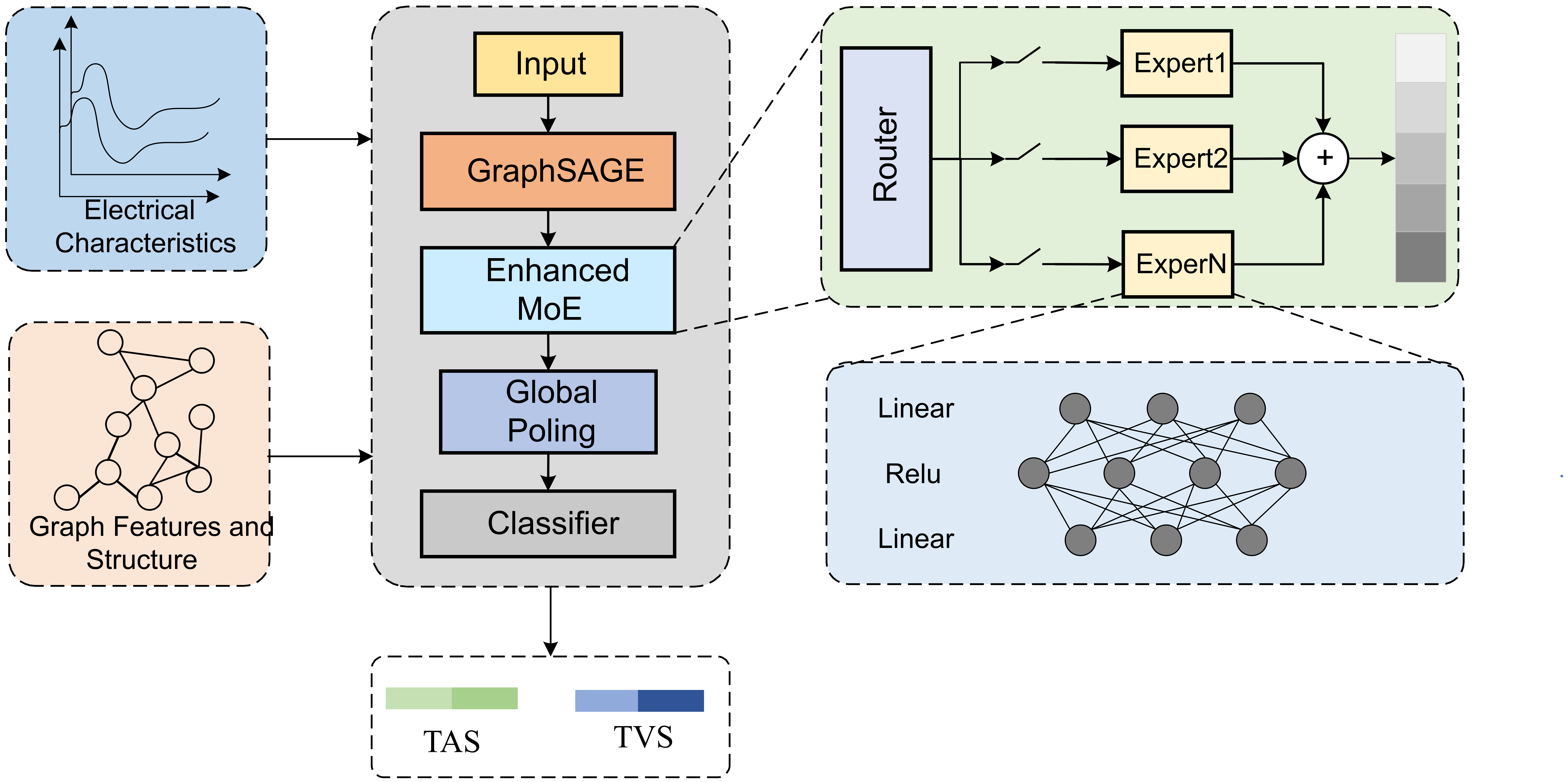}
  \caption{Architecture of the MoE-GraphSAGE Model}
  \label{fig:architecture}
\end{figure}

The architecture of the proposed MoE-GraphSAGE model is illustrated in Fig.~\ref{fig:architecture}.

\subsection{GraphSAGE Architecture}

The power system topology is naturally represented as a non-Euclidean graph, with nodes as generators, buses, or loads, and edges as transmission lines. GraphSAGE is employed to capture the spatiotemporal features of system disturbances:
\begin{equation}
H = \text{GraphSAGE}(X, A)
\end{equation}
where $X$ is the node feature matrix and $A$ stands for the system adjacency matrix.

\subsection{Mixed Expert (MoE) Mechanism}

The core of the MoE mechanism is the gating network, which adaptively assigns weights to each expert according to the input features~\cite{zhang2025mvho}. Specifically, the gating network:
\begin{equation}
g = \text{Gate}(h)
\end{equation}
Where $g$ denotes the expert weight vector, Gate represents the gated network, and $h$ refers to the input features.

Each expert's output $E_i$ is combined via weighted aggregation to produce the final output:
\begin{equation}
y = \sum_{i=1}^{N} g_i E_i(h)
\end{equation}

\section{Procedure for Integrated Power Angle and Voltage Stability Assessment}

\begin{figure}[htbp]
  \centering
  \includegraphics[width=0.9\linewidth]{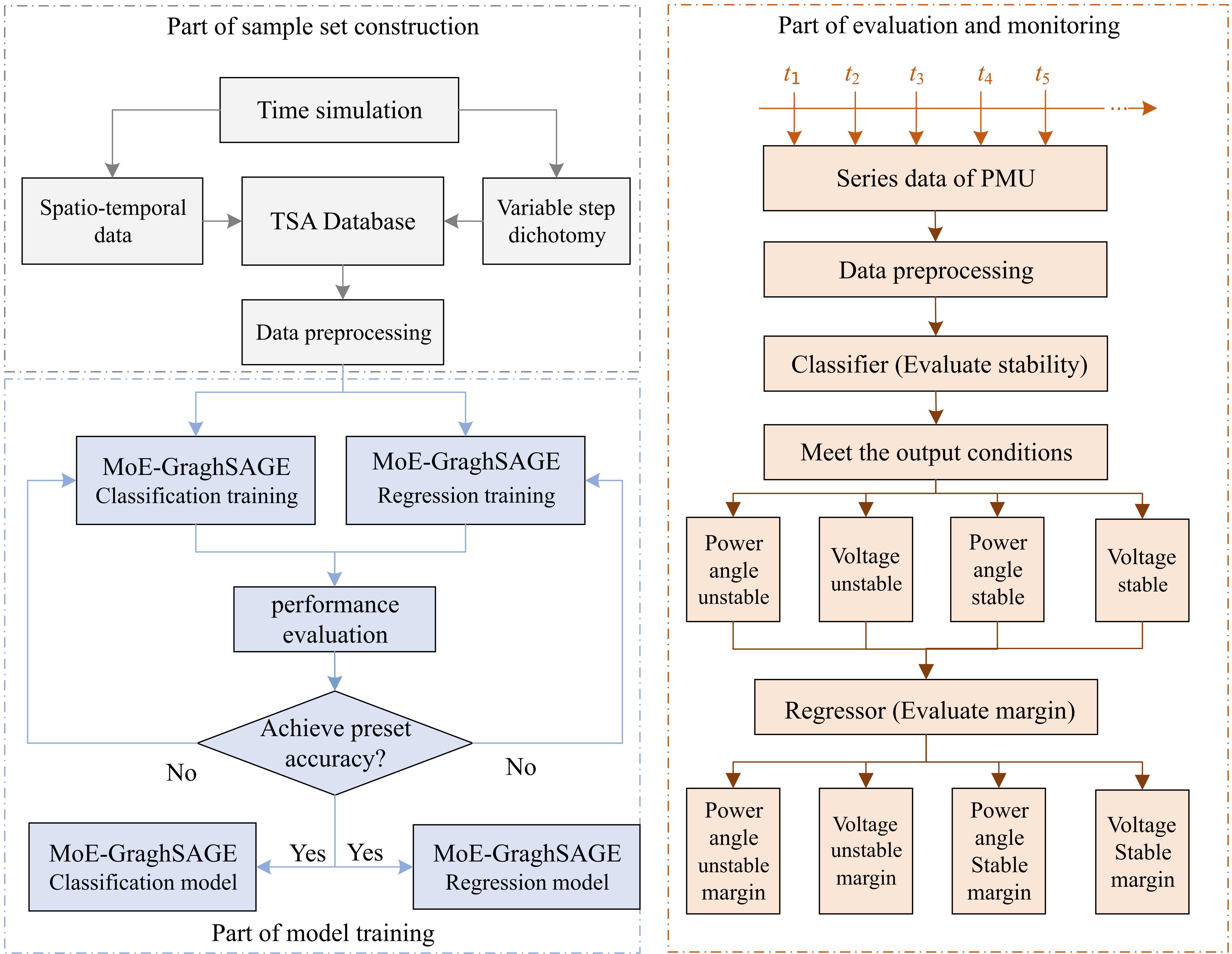}
  \caption{TVS\&TAS integrated assessment framework}
  \label{fig:framework}
\end{figure}

Figure~\ref{fig:framework} illustrates that the comprehensive evaluation of transient angle stability (TAS) and transient voltage stability (TVS) involves three stages: sample dataset construction, model training, and online monitoring. This section provides a brief overview of these stages.

\subsection{Part of Sample Set Construction}

A complete sample set consists of features and labels. In this study, the construction process is as follows:

\textbf{Step 1:} Feature construction. The time-varying power system topology is used as spatial features, while node voltages and phase angles serve as temporal features capturing transient characteristics. Batch simulations generate the corresponding data, yielding a spatiotemporal feature set for model training~\cite{shi2023b}.

\textbf{Step 2:} Labeling TAS and TVS. By evaluating the stability criteria for the given operating conditions, the system's stability or instability is identified, and binary classification labels for rotor angle and voltage are subsequently derived.

\textbf{Step 3:} Margin labeling of TAS and TVS. The time boundaries of TAS and TVS are identified using a variable-step dichotomy method, and safety margin labels are obtained based on critical clearing times.

\subsection{Part of Model Training}

In the training stage, both classification and regression tasks are performed. The classification task identifies the steady-state conditions of power angle and voltage under disturbances, while the regression task quantifies the stability margin or degree of instability. These tasks are trained concurrently, and training is terminated when the model achieves a predefined accuracy threshold, ultimately yielding both classification and regression models.

\subsection{Part of Online Monitoring}

During online monitoring, the comprehensive model processes real-time PMU time-series data. First, the classification model evaluates transient stability and outputs a stable or unstable decision. Then, the regression model quantifies the stability level or instability margin, enabling a detailed assessment of power system stability.

\section{Stability Criteria and Model Evaluation Metrics}

\subsection{Criterion for TAS and TVS}

\subsubsection{Criterion for TAS}

In this paper, sample classification is performed using the transient stability index (TSI):
\begin{equation}
\text{TSI} = \delta_{\max} - \delta_{\min}
\end{equation}
Where $\delta_{\max} - \delta_{\min}$ is the maximum power angle difference between any two generators during operation. When $\text{TSI} < 180°$, the system remains stable under this disturbance condition; otherwise, it indicates system instability.

\subsubsection{Criterion for TVS}

This paper employs a practical criterion based on a single binary table to describe and evaluate transient voltage stability:
\begin{equation}
\text{TVS} =
\begin{cases}
\text{stable}, & \text{if } V_{\min} > V_{\text{th}} \text{ and } t < t_{\max} \\
\text{unstable}, & \text{otherwise}
\end{cases}
\end{equation}
Where $V_{\min}$ is the minimum post-disturbance node voltage, $V_{\text{th}}$ denotes the specified voltage threshold, $t$ is the actual duration that voltage stays below this threshold, and $t_{\max}$ represents the maximum allowable duration. A value of $\text{TVS} = \text{stable}$ indicates transient voltage stability, otherwise instability. The parameter settings in this work are based on practical engineering criteria~\cite{dl2013}, which define the system as voltage-stable if the load bus voltage restores to 0.8 pu within 10 seconds following a fault.

\subsubsection{Stability Degree Metrics}

This paper introduces a safety margin index based on the critical clearing time (CCT) for TVS and TAS, characterizing the system's real-time state using stability margin and instability degree. The stability margin and instability degree indicators are normalized and standardized as follows:
\begin{equation}
M = \frac{t_{\text{CCT}} - t_{\text{pred}}}{t_{\text{CCT}}}
\end{equation}
\begin{equation}
D = \frac{t_{\text{pred}} - t_{\text{CCT}}}{t_{\text{pred}}}
\end{equation}
\begin{equation}
\bar{M} = \frac{1}{N}\sum_{i=1}^{N} M_i
\end{equation}
Where $t_{\text{CCT}}$ denotes the safety margin indicator for both TVS and TAS, $t_{\text{pred}}$ is the fault clearing time, and $M$ is the model-predicted output time. $M$ represents the normalized safety margin, where $D$ indicates the degree of instability and $M$ indicates the stability margin. $\bar{M}$ is the normalized margin or instability index, and $N$ is the number of samples.

\subsection{Model Evaluation}

\subsubsection{Evaluation Indicators of Classification Model}

Accordingly, this paper introduces four metrics derived from the confusion matrix to comprehensively assess the performance of transient stability evaluation. Based on the confusion matrix in Table~\ref{tab:confusion}, accuracy $A$, missed detection rate $MDR$ (the ratio of unstable samples classified as stable), false positive rate $FPR$ (the ratio of stable samples classified as unstable), and geometric mean $G$ are defined.

\begin{table}[htbp]
\caption{Two dimensional confusion matrix}
\label{tab:confusion}
\centering
\begin{tabular}{ccc}
\toprule
Sample & Predicted Stable ($\hat{y}_0$) & Predicted Unstable ($\hat{y}_1$) \\
\midrule
Actual Stable ($y_0$) & $n_{00}$ & $n_{01}$ \\
Actual Unstable ($y_1$) & $n_{10}$ & $n_{11}$ \\
\bottomrule
\end{tabular}
\end{table}

In Table~\ref{tab:confusion}, $y_0$ and $y_1$ correspond to the true stable and unstable labels, $\hat{y}_0$ and $\hat{y}_1$ correspond to the predicted stable and unstable labels, and $n_{ij}$ represents the number of samples at row $i$ and column $j$. The integrated classification assessment indicators are described based on the confusion matrix as follows:
\begin{equation}
A = \frac{n_{00} + n_{11}}{n_{00} + n_{01} + n_{10} + n_{11}}
\end{equation}
\begin{equation}
\text{MDR} = \frac{n_{10}}{n_{10} + n_{11}}
\end{equation}
\begin{equation}
\text{FPR} = \frac{n_{01}}{n_{00} + n_{01}}
\end{equation}
\begin{equation}
G = \sqrt{\frac{n_{00}}{n_{00} + n_{01}} \times \frac{n_{11}}{n_{10} + n_{11}}}
\end{equation}

\subsubsection{Evaluation Indicators of Regression Model}

This paper selects Mean Squared Error (MSE) and Mean Absolute Error (MAE) as evaluation metrics for regression models.

\section{The Example Analysis}

\subsection{Construction of Sample Set}

The sample generation is performed on the improved New England 39-bus system, accounting for the dynamic characteristics of both loads and generators under diverse operating conditions. The load model adopts a composite structure of induction motors and constant impedance, with motor shares of 50\%, 60\%, and 70\%. On the system's 34 transformerless AC lines, severe three-phase short-circuit faults are simulated at 10\%, 30\%, 50\%, 70\%, and 90\% of the line lengths. In the simulation, each fault initiates at 1s, lasts 10s, and is sampled at 0.01s, with clearance times ranging from 3 to 11 cycles. This setup yields $34 \times 5 \times 3 \times 9 = 4590$ sample features. Based on transient angle and voltage stability assessments, 106,920 classification labels are obtained. The regression model uses the same training features as the classification model. For labeling, TVS\&TAS were annotated based on safety margin metrics, generating 106,920 safety margin labels for the regression model.

\subsection{Analysis of Evaluation Results}

\subsubsection{Comparison of the Performance of Various Algorithms}

To evaluate the effectiveness of the proposed MoE-GraphSAGE model for transient stability assessment, it is compared with standard graph neural networks, including GAT~\cite{velickovic2018}, GCN~\cite{sharma2022}, GIN~\cite{steiger1992}, and GMT~\cite{chen2025}, as well as hybrid models HL-GNN~\cite{liu2024} and TG-GNN~\cite{zhou2024}. The dataset is split into training (70\%), validation (10\%), and testing (20\%) subsets, and each model is trained 10 times to reduce stochastic variability. The test set assessment results are presented in Table~\ref{tab:performance}.

\begin{table}[htbp]
\caption{Performance Comparison of Different Algorithms}
\label{tab:performance}
\centering
\begin{tabular}{lcccc}
\toprule
Model & $A$ (\%) & $MDR$ (\%) & $FPR$ (\%) & $G$ (\%) \\
\midrule
GAT & 95.38 & 95.82 & 95.12 & 95.43 \\
GCN & 94.92 & 95.38 & 94.72 & 95.07 \\
GIN & 95.62 & 95.92 & 95.02 & 95.47 \\
GMT & 95.22 & 95.62 & 94.92 & 95.23 \\
HL-GIN & 98.12 & 95.82 & 95.22 & 95.48 \\
TG-GNN & 97.92 & 95.62 & 94.82 & 95.18 \\
\textbf{Proposed} & \textbf{98.67} & \textbf{96.70} & \textbf{99.62} & \textbf{98.15} \\
\bottomrule
\end{tabular}
\end{table}

In contrast, MoE-GraphSAGE achieves the best overall performance, with 98.67\% accuracy and 99.62\% recall—nearly 4\% higher than hybrid models—substantially lowering missed instability detections. These results indicate that integrating the MoE mechanism with GraphSAGE enables adaptive recognition of diverse instability modes and comprehensive extraction of spatiotemporal features, significantly outperforming existing approaches in both accuracy and reliability.

\subsubsection{Comparison of the Computational Efficiency Among Different Algorithms}

\begin{figure}[htbp]
  \centering
  \includegraphics[width=0.65\linewidth]{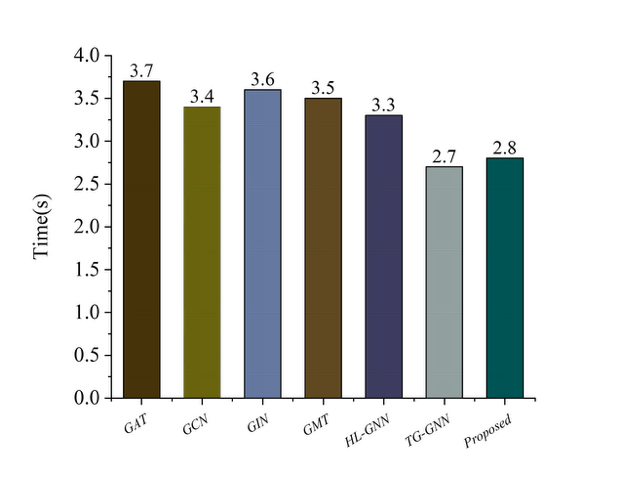}
  \caption{Computational efficiency of different algorithms}
  \label{fig:efficiency}
\end{figure}

To assess the computational efficiency of the models in practical applications, the runtime of each model on the test set was tested on a single NVIDIA A100 40G GPU (refer to Fig.~\ref{fig:efficiency}).

The results indicate that traditional models, including GAT, GCN, and GIN, run in about 3.5s. GMT and HL-GNN show moderately increased runtimes, while TG-GNN and the proposed MoE-GraphSAGE achieve runtimes below 3s. Relative to the benchmark methods, MoE-GraphSAGE achieves high computational efficiency without compromising accuracy, highlighting its promising real-world applicability.

\subsubsection{Assessment of Security Margins}

The MoE-GraphSAGE regression model is trained using the power angle and voltage stability margin labels generated in Section 4.2.1. In the test set, the MSE and MAE for transient voltage are 0.00017 and 0.0091, respectively; for transient rotor angle, MSE is 0.00023 and MAE is 0.0100. The results demonstrate that MoE-GraphSAGE accurately assesses the stability levels or instability margins associated with stable conditions, achieving superior performance in multi-task transient stability margin evaluation.

\section{Conclusion}

This study presents a transient stability evaluation model based on gated MoE-GraphSAGE, verified on the improved IEEE 39-bus system. The model efficiently extracts spatiotemporal topological features, adaptively identifies instability modes, and achieves 98.67\% accuracy under a multi-task framework. Its gated mixture-of-experts mechanism enhances adaptability to diverse operating conditions, while maintaining computational efficiency suitable for real-time deployment, demonstrating practical feasibility for online stability assessment. Future work will explore extending this framework to other power system applications, such as renewable energy integration and intelligent grid management~\cite{shi2025multi}.

\section*{Funding}
The work was supported in part by National Natural Science Foundation of China [Grants Nos. 52037006].

\bibliographystyle{unsrt}

\end{document}